\documentclass[11pt,preprint]{aastex}

\usepackage {pstricks}

\shorttitle{Core-Collapse Progenitors}
\shortauthors{Arnett \& Meakin}

\newcommand{\etal} { et~al.\ }  
\newcommand{\sol}{$M_\odot$}
\def \nuc#1#2{\relax\ifmmode{}^{#1}{\protect\text{#2}}\else${}^{#1}$#2\fi}
\def \msol#1{\relax$#1\,M_\odot\/$}

\begin{document}

\title{Toward Realistic Progenitors of Core-Collapse Supernovae}

\author{W. David Arnett\altaffilmark{1,2}
\email{ wdarnett@gmail.com}
}
\author{Casey Meakin\altaffilmark{1,3}
\email{casey.meakin@gmail.com }
}
\altaffiltext{1}{Steward Observatory, University of Arizona, 
933 N. Cherry Avenue, Tucson AZ 85721}

\altaffiltext{2}{ICRAnet, Rome, Pescara, Nice}

\altaffiltext{3}{Theoretical Division, Los Alamos National Lab,
Los Alamos, NM}

\begin{abstract}

Two-dimensional (2D) hydrodynamical simulations of progenitor evolution  
of a 23 \sol \ star,  close to core collapse (in $\sim1$ hour in 1D), with simultaneously active C, Ne, O, and Si burning
shells, are presented and contrasted to existing 1D models (which are forced to
be quasi-static). Pronounced
asymmetries, and strong dynamical interactions between shells are seen in 2D.
Although instigated by turbulence, the dynamic behavior proceeds to sufficiently
large amplitudes that it couples to the nuclear burning. Dramatic growth of
low order modes is seen, as well as large deviations from spherical symmetry in the burning shells.
The vigorous dynamics is more violent than that seen in earlier burning stages in the 3D simulations of 
a {\em single} cell in the oxygen burning shell \citep{ma07b}, or in 2D simulations not including an active Si shell. 
Linear perturbative analysis does not capture the chaotic behavior of turbulence
(e.g., strange attractors such as that discovered by \cite{lorenz}),
and therefore badly underestimates the vigor of the instability.

The limitations of 1D and 2D models are discussed in detail.
The 2D models, although flawed geometrically, represent a more realistic
treatment of the relevant {\em dynamics} than existing 1D models, and present a
dramatically different view of the stages of evolution prior to collapse. 
Implications for interpretation of SN1987A, abundances in young supernova remnants, pre-collapse outbursts, progenitor structure, neutron star kicks, and fallback are outlined. 
While 2D simulations provide new qualitative insight, fully 3D simulations are needed for a quantitative understanding of this stage of stellar evolution. The necessary properties of such simulations are delineated.

\end{abstract}

\keywords{stars: evolution,  hydrodynamics: convection -turbulence,  supernova:
-remnants -core-collapse -nucleosynthesis}

\section{Introduction}

The first detailed calculations of the final, neutrino-cooled burning stages prior to core collapse
of massive stars were done by \cite{rsz67},
with simplified energy generation rates, and by \cite{wda68} using a small (24 species)
network.
Because of the extreme demands upon computer resources then available,  simplified
energy generation rates were used in subsequent calculations; see \cite{wda96}
for references to the early work.
This work showed two issues which have not yet be resolved: (1) the definition of convective zone boundaries involves incomplete physics (by construction 
mixing length theory ignores gradients, \cite{spie71}), and (2) the stability of the burning in a convective region has not been demonstrated, only assumed. 
While thermal instability  (thermal {\em stability} implies a global balance between nuclear heating and neutrino cooling in the convective zone) was considered \citep{wda72a,wda96}), dynamical instability (i.e., instability related to fluid flow) was not.
In this paper we will show that both these issues are significant, 
and that they can be resolved by 3D numerical simulations and theory.

Almost all previous progenitor models for core collapse have focused on thermal
behavior and quasi-static mixing, which are described by the evolution of
the temperature and the composition variables. The dynamic behavior of the
stellar plasma includes and is dominated by the velocity fields, which not only imply mixing, but also possible change in the stellar structure. The star is not necessarily a quasi-static object, but may have significant fluctuations in its variables. The dynamical behavior, found here for simultaneous, multi-shell
burning, will drive entrainment \citep{ma07b} at convective shell boundaries,
changing the nucleosynthesis yields and the size of the Fe core at collapse.

The presupernova structure of a massive star consists of a core, mantle, and envelope. The envelope is extended, composed of H and He, and may have been removed prior to core collapse by wind-driven mass loss, or tidal stripping by a companion. The mantle is composed of burning shells of C, Ne, O and Si; these shells are convective, interact nonlinearly, contain most of the nucleosynthesis products ejected, and may smother a neutrino-driven explosion. The core is composed of Fe-peak nuclei, and its mass is determined by its entropy and electron fraction. Lower entropy and electron fraction give smaller cores, which are easier to explode by neutrino transport mechanisms. Core collapse mechanisms for explosion are sensitive to core mass, mantle density, and rotation; all are sensitive to the treatment of turbulence.
The simulations described here involve the simultaneous action of C, Ne,
O and Si burning shells.
The oxygen shell is special, because (1) formation of electron-positron pairs
softens the equation of state, aiding the formation of large amplitude motion, 
(2) the large abundance of oxygen and its relatively large energy release per
unit mass provide a large thermonuclear energy reservoir, and (3) oxygen burning,
unlike silicon burning,
has little negative feedback from quasi-equilibrium to damp flashes (see below).
 
In Section~2 we discuss the historical context of progenitor models of core collapse
supernovae, focusing on issues of mixing,  causes of time-dependence and multi-dimensionality, prospects for development of better computational tools,
and the differences between 2D and 3D simulations.
In Section~3  we describe our 2D simulations of progenitor evolution with multiple,
simultaneously active, burning shells (C, Ne, O, Si), and discuss some new phenomena which appear.
In Section~4 we summarize the implications for evolution prior to core collapse.
In Section~5 we focus on several problems in observational astrophysics which need to be reconsidered in light of our results.
In Section~6 we summarize the major conclusions, and outline the necessary features of new 3D simulations which will be required to quantitatively resolve the issues presented by the 2D simulations.

\section{Brief Historical review of 1D, 2D, and 3D Models}

\subsection{1D Models: Mixing}

By the term {\em advection} we mean the transport of a parcel of matter by
a large-scale flow; by {\em diffusion} we mean the transport of a parcel of matter
by a random walk of small-scale motions (often microscopic motion of ions
in the stellar plasma).
The mass (baryon) flux due to advection is
\begin{equation}
{\bf F_m} = {\bf u} \rho,
\end{equation}
where $\bf u$ is the fluid velocity vector and $\rho$ is the mass (baryon) density.
The flux of any scalar variable is related to this flux by a factor of the 
density of the variable per unit mass (baryon). For example, if the mass (baryon)
fraction of nuclear species $i$ is $X_i$, the flux of this species is 
\begin{equation}
{\bf F_i} = {\bf u} \rho X_i = {\bf F_m} X_i.
\end{equation}
The rate of change due to such fluxes involves a divergence, as in the continuity equation, 
\begin{equation}
\partial \rho / \partial t = - {\bf \nabla \cdot F_m}.
\end{equation}
Thus, 1D advection involves a first order spatial differential operator.

If the considered volume contains heterogeneous (turbulent) matter, 
the flux may vary over
the corresponding surface. If that volume is a zone in a stellar evolution computation,
then the fluxes must be averaged over the heterogeneity, and some knowledge
(or assumption) regarding the smaller scale structure is required. Consider the
simple example of only an inflow and an outflow, and variation only in the
vertical direction. The net flux for species $X_i$ is then
\begin{equation}
F_{net}(X_i) = - (F_m X_i )_{out} + (F_m X_i)_{in}.
\end{equation}
Suppose that $F_{out} =  F_{in} \equiv \rho  u$,  so $F_{out}-F_{in}= 0$,
which automatically satisfies
mass (baryon) conservation, so that
\begin{equation}
F_{net}(X_i) = -\rho u \Big [ (X_i)_{out} - (X_i)_{in} \Big].  \label{advectx}
\end{equation}
In this case the flux (of composition $i$) is proportional to the negative of the
difference in abundance in the up and down flows\footnote{A more realistic
and relevant case for stellar turbulent convecction would allow different speeds in up and down flows; that complication is not necessary here.}.

For diffusion, the number flux follows from Fick's law,
\begin{equation}
{\bf \mathcal F_d}(X_i) = - {1\over3} \lambda v {\bf \nabla} N_i, 
\end{equation}
and is proportional to the gradient of a number density $N_i$. Here $\lambda$ is
a mean-free-path and $v$ is a speed of diffusing particles.
The number density for species $i$ may be written as
$N_i = \rho {\cal A} X_i/A_i$, where ${\cal A} = 1/m_{amu}$ is Avogadro's number (the inverse of the atomic mass unit) and
$A_i$ is the number of nucleons in species $i$. 
The mass (baryon) flux is ${\bf F_d}(X_i) = A_i m_{amu} {\bf \mathcal F_d}(X_i) $.
Thus,
\begin{equation}
{\bf F_d}(X_i)  = -{1 \over 3} \lambda v  {\bf \nabla} (\rho X_i). \label{diffusionx}
\end{equation}
If we consider the divergence of the flux, diffusion implies a second order spatial differential operator, in contrast to advection which, as we saw, implies first order. This is a fundamental mathematical difference. 

\cite{wzw78,whw02} introduce an effective diffusion coefficient $D$ to model convection:
\begin{equation}
{d Y_i  \over dt} = {\partial \over \partial m}\Big [ (4 \pi r^2)^2 \rho^2 D {\partial Y_i \over \partial m } \Big ].
\end{equation}
In the case of a region unstable to convection by the Ledoux criterion, they take
$D_c \equiv v_c \ell/3$, where $v_c$ is the velocity estimated by mixing-length
theory \citep{bv58,clay83} and $\ell$ is the mixing length. Thus if 
\begin{equation}
\lambda v = \ell v_c, \label{wzwapprox}
\end{equation} 
then Equation~ \ref{diffusionx} is recovered.
The use of Equation~\ref{diffusionx} requires that $ \lambda \ll \Delta r$, 
where $\Delta r$ is a zone size.
For a turbulent cascade, $\ell \gg \Delta r$, which is inconsistent with the previous
requirement.
The treatment of  convection as diffusion is essentially an algorithmic interpolation
procedure, but has 
an intrinsic contradiction in physics. This arises because turbulent flow has two
facets: a flow at  large scale $\ell$ which does most of the transport, and a flow at  small scale  $\lambda$ which does the dissipation, and $\ell \gg \lambda$.  The diffusion
approximation might be used for the small scale flow, but that is irrelevant to the
transport problem, which is dominated by the large scale flow \citep{amy09}.

Notice the rough similarity between Equation~\ref{advectx}, which represents the underlying fluid dynamics \citep{ll59}, and Equation~\ref{diffusionx}, which represents the proposed approximation.
Taking the density outside the differential operator and  using a finite difference
representation of Equation~\ref{diffusionx}, 
\begin{equation}
{\bf F_d(X_i) } \sim -\Big [ {1 \over 3} \lambda v \rho / \Delta r \Big ]  \Big [ (X_i)_{out}- (X_i)_{in} \Big ]. 
\end{equation}
For this to be consistent with 1D advection (Equation \ref{advectx}) we must have
a diffusion rate that is dependent upon the zoning! While perhaps useful
algorithmically, this is not clear conceptually; see also the discussion of 
\cite{mm00}, who also express doubts concerning the approximation of advection by diffusion.

Mathematically, diffusion is a procedure which maximally smooths
gradients, so that a more realistic procedure may be expected to exhibit
less smoothing. 
While based on poor physics, the diffusion model of convection had the virtue that it allowed numerical prediction of yields.

Much effort has recently gone into extending stellar evolution to include rotational
effects \citep{zahn92,cdp95,mm00,tas,whw02}. Evolution of a rotating star will develop differential rotation in general, and shear flow. Similarly, deceleration of
convective plumes will develop shear flow. In stars both flows will be turbulent. Convection and rotation have an underlying similarity not reflected in stellar evolution theory.
The review of \cite{mm00} gives a clear presentation of the various approximations
involved in reducing the full fluid dynamic equations to a simpler, more easily solvable set. It is now possible to test these approximations by numerical
simulation in both shearing box \citep{am09iau, sg10} and whole star domains
\citep{bp09,bbbmt10}; see also the theoretical developments of \cite{balbus09,bw10}. A theoretically sound approach must treat the shear from
differential rotation and the shear from convective plumes on a consistent basis,
reflecting their underlying physical similarity \citep{turner73}.

  \subsection{Time Variation\label{time-dep}}
  \noindent{\bf The $\epsilon$-mechanism.}
The $\epsilon$-mechanism \citep{ledoux41,ledoux58} for driving stellar pulsations
by nuclear burning was examined  by \cite{wda77} for Si burning in 1D geometry,
using the simple energy generation rate proposed by \cite{bcf68}. In this case,
explicit 1D hydrodynamic simulations gave nuclear-energized pulsations (see
Figure~6 in \cite{wda77}). There were two high frequency modes
(period ~$0.1$ seconds and $1$ second) and a lower frequency convective mode 
(turnover time $\sim20$ seconds). The intermediate frequency mode was
related to the Si flash.
Computation with a realistic nuclear network subsequently showed that the highest frequency (``acoustic") pulsations would be strongly damped due to the 
quasi-equilibrium nature of Si burning; an increase in temperature gave an increase
in free alpha particles (and neutrons and protons), which required energy, and
resisted the increase in temperature (with an almost 180 degree phase shift, meaning
that there is strong negative feedback to resist changes).
However, this damping process does not apply to O, Ne or C burning, which 
have little or no quasi-equilibrium behavior, and in principle could drive pulsations
more vigorously.

\noindent{\bf The $\tau$-mechanism.}
\cite{am11b}, in study of the 3D simulations of \cite{ma07b}, found that the bursts
in turbulent kinetic energy were related to similar fluctuations in the \cite{lorenz}
 model of a convective roll; this is the now-famous strange attractor.
This instability mechanism, called the $\tau$-mechanism for ``turbulence", is
expected to be a general property of stellar convection zones, and probably
is the cause of the fluctuations in luminosity observed in irregular variables
(e.g., Betelgeuse,  see \cite{am11a,am11b}).
This process is independent of the temperature and density dependence of the
thermonuclear heating, and thus is distinctly different from the $\epsilon$-mechanism.
Unlike the $\epsilon$-mechanism, the $\tau$-mechanism is not a linear instability, but is inherently nonlinear. 

In appropriate circumstances however, the $\tau$-mechanism may couple to nuclear burning, making pulsations more violent, giving a more complex, combined $\epsilon+ \tau$ mechanism.
The $\tau$ mechanism involves a {\em non-linear} instability, unlike the linear
instabilities discussed by \cite{gold97}. 
In a detailed analysis, \cite{mbh06} solve the {\em linear} perturbation
equations for several massive stars prior to core collapse, and based upon
the $\epsilon$ mechanism alone, find inadequate driving to cause such violent behavior as is actually shown in the non-linear simulations presented below.

The analysis of  \cite{mbh06} is perfectly correct within framework of the linear assumption, but the assumption fails.
The linear approximation is not valid in this case because turbulence is not a linear
perturbation of the system. \cite{lorenz} showed that chaotic behavior in a convective roll is due to the {\em nonlinear} interaction between temperature gradients (both vertical and horizontal) and convective speed. Solutions which are initially close to each other will diverge exponentially as time passes. 
Thus the interplay between pulsation and
turbulent convection is not captured by traditional linear perturbation analysis
of stellar pulsations. This is an interesting theoretical result, suggesting
that linear perturbative methods for pulsations  \citep{cox80,unno89} may require reexamination when turbulent convection is important (as those authors feared).

 \subsection{History of Multi-dimensional Progenitor Models}
 
 There have been relatively few multi-dimensional simulations of core collapse
 progenitors, but there has been a rich context of efforts on turbulence and
 stellar hydrodynamics  \citep{pw94,pw00,fma89}, on the helium flash
 \citep{deupree84,deupree96,deupree03,dearborn06,mocak08,mocak09,mocak10}, 
 and on turbulent MHD with rotation (e.g., m-dwarf simulations by
 \cite{browning08}), as well as extensive work on the Sun with the ASH code
 (e.g., \cite{bbbmt10})
 and on stellar atmospheres pioneered by Nordlund and Stein (e.g., \cite{nsa}). This context
 has speeded the development of tools and helped determine their reliability.
 The first effort at simulating oxygen burning \citep{wda94} helped define the shell burning problem with regard to required resolution, but suffered from the use of sectors so narrow in angle that boundary effects affected the flow.

\begin{deluxetable}{lrrrrr}
\tablecaption{Two-dimensional Progenitor Simulations\label{table1}}
\tabletypesize{\small}
\tablewidth{390pt}
\tablehead{ \colhead{Reference} &
\colhead{$a$} & \colhead{$b$}  & \colhead{$c$} & \colhead{$d$} & \colhead{$e$}
}
\startdata
zoning & 256x64 & 256x64 & 172x60  & 800x320 & 800x320 \\
code  & PROMETHEUS  & PROMETHEUS   & VULCAN     & PROMPI & PROMPI \\
eos$\rm^f$ & wda & wda & wda & TS & TS \\
network  & 12 & 123 & 12 & 25 & 37 \\
burning & O &  Si &O & C,Ne,O & C,Ne,O,Si \\
core & Si & Si & Si & Si & Fe\\ 
duration(s) & 300 & ~200 &  1200 & 2500 & 600 \\
\enddata
\tablenotetext{a}{\cite{ba98,ba94}}
\tablenotetext{b}{\cite{ba97b}, a small inert spherical boundary surrounded by Si.}
\tablenotetext{c}{\cite{aa00}}
\tablenotetext{d}{\cite{ma06}, energy release verified against a 177 nuclei network}
\tablenotetext{e}{\cite{meakin06} and this paper, energy release verified against a 177 nuclei network}
\tablenotetext{f}{See \cite{ta99} and
\cite{ts00}.}
\end{deluxetable}

\placetable{1}

Table~\ref{table1} summarizes aspects of early 2D simulations of progenitor models
in comparison to the present work.
 The first 2D hydrodynamic simulations of core-collapse progenitors (oxygen burning) showed striking new phenomena: mixing beyond formally stable boundaries, hot spots of burning due to $\rm C^{12}$ entrainment, and inhomogeneity in neutron excess \citep{ba94}. Further work \citep{ba98}
confirmed that convection was too dynamic to be well represented by diffusion-like
 algorithms,  that large density perturbations (8\%) formed at convective
 boundaries, and that gravity waves were vigorously generated by the flow.
 Extension to Si burning \citep{ba97b} with a 123 
isotope network showed similar highly dynamic
behavior and significant inhomogeneity in  neutron excess.
With an entirely different 2D code (VULCAN), \cite{aa00} extended the evolution of
the oxygen shell of \cite{ba98} to later times, and discovered that the extensive
wave generation at the convective boundaries induced a slow mixing in the
bounding non-convective regions. 

\cite{kwg03} investigated shell oxygen burning in 3D with an anelastic hydrodynamics code \citep{gg84}, and found small density and pressure perturbations (less than 1\%). The boundary conditions were impermeable and stress-free, so that convective overshoot could not be studied. They concluded
that, contrary to previous work listed above (done with 2D compressible codes),
the behavior was not very dynamic and could be described by the MLT algorithms
used in 1D stellar evolution codes (e.g., \cite{wzw78}). \cite{ma07a}, using 3D compressible hydrodynamics, showed that the differences were due to the choice of (unrealistic) boundary conditions that \cite{kwg03} used. Inside the convection zone,
away from the boundaries, the Glatzmaier code gave results in good agreement with
the compressible code. However, as stressed in \cite{ba98},
fluid boundaries allow surface waves to build to large amplitudes ($\delta \rho / \rho \sim 0.1$), so that the hard boundaries used in \cite{kwg03} distorted the physics. The good agreement between the anelastic and the compressible solution {\em within} the convection zone, and the agreement between the stable layer dynamics given by the compressible fluid code and
analytic solutions to the non-radial wave equation, indicate that the compressible
hydrodynamic techniques are robust for this problem, even for Mach numbers below 0.01. An anelastic code with the correct boundary conditions should give the  same result; the flow is subsonic.

With the PROMPI code \citep{meakin06}, several new series of computations were done. \cite{ma07a,ma07b} presented the first 3D calculations of the phase of shell
O burning with full physics (i.e., compressibility, nuclear network, real equation of state, appropriate boundaries, etc.). 
\cite{ma06} calculated in 2D the oxygen burning shell, and both C and
Ne burning shells above this.  
Here we present simulations \citep{meakin06} with similar microphysics, extended to include the silicon burning shell as well, so that C, Ne, O and Si burning occur on the grid, but only in 2D.

\subsection{Future Prospects: Beyond 2D}

Progress will not be a simple progression reflecting the growth of computational
resources, but also of theoretical understanding.
\cite{am11b} have shown a connection between the Lorenz model of a convective
roll and the bursty pulses in turbulent kinetic energy seen originally in the
\cite{ma07b} simulations, which seem to have a similar strange attractor.
The Lorenz model is a low order (3 variable) dynamical system, and its physical
identification with the 3D simulations suggests how we may project the essence
of the 3D solutions onto 1D for stellar evolution and dynamics (``321D").
This will give an immediate improvement over MLT as well as better initial models
for numerical simulation; see below. The physical basis is strengthened
mathematically by use of the Karhunen-Lo\`eve or proper orthogonal decomposition (POD; see \cite{hlb96})
of the 3D numerical data set of \cite{ma07b}. Preliminary results show that roughly half of the turbulent kinetic energy is in the single lowest order empirical eigenmode, 
supporting the idea that low order dynamical systems may be used to describe the complexities of time dependent turbulent flow in stellar convection zones.

The dynamical phases prior to core collapse may not attain a statistically
averaged state, so that these methods (321D and KL decomposition), while
promising for earlier evolution, may not be the optimum tools for the pre-collapse
itself, where large eddy simulations (ILES, \cite{boris}) in 3D are needed. 
However, use of the theoretical tools (321D and KL decomposition) provides
a means of estimating the range of fluctuations about an given ILES simulation,
which may be tested insofar as computational resources allow, for other initial
conditions.

 \subsection{Comparison of 2D and 3D Simulations}

\cite{ma07b} show a precise comparison of two simulations, which use the same microphysics,  initial model and code,
and differ only in dimensionality, which changes from 2D to 3D. The simulations are
  discussed there as ``core convection" (see Figure~4 therein).
 The topology of the convective flow is significantly different
 between 2D and 3D models: the 3D convective flow is dominated by small plumes and eddies while the 2D flow is much more laminar, and dominated by large vortices
 (``cyclones") which span the depth of the layer.  
 The 2D vortices trap material which is slow to mix with surrounding matter; 
 in 3D the flows become unstable
 and matter mixes more completely. The wave motions in the stable layers do
 not have an identical morphology in 2D and 3D, and the velocity amplitudes are much larger in 2D.
 
 The differing behavior is due to the constraint of geometry, and the law of angular momentum conservation, which forces the formation of large cyclonic patterns in 2D that are unstable  in 3D. The turbulent cascade moves from small scale to large (cyclones) in 2D, but from
 large to small (Kolmogorov cascade) in 3D.
  Cyclonic behavior at large scales is physically reasonable for the Earth's
 atmosphere, for example, which because of its restriction in height, is approximately  two-dimensional\footnote{The density scale height in the vertical direction is small compared to the width of a typical cyclonic system seen on the evening news. Oxygen would be required in an unpressurized aircraft at 35,000 feet, which is a measure of the ``height" of the atmosphere. This is less that one percent of the width of a large cyclonic system. This is a very flat domain, and with rotation, is dominated by geostrophic (i.e., 2D) flow patterns at the large scales.}.  In stars there is no physical constraint to enforce 2D motion, so that 2D simulations of stars are not realistic, merely computationally less expensive.
 
 In \cite{ma07b},  3D simulations were done for a single cell in the O shell; the computational demands for simultaneous multi-shell burning are more extreme
 (but almost feasible). As a first step we present 2D simulations, which though
 flawed, are instructive. 
 We describe the first extended results for four simultaneous burning shells (Si burning included, \cite{meakin06}), and interpret them with the benefit of extensive analysis of 3D O burning results.
Taken with Figure~4 from \cite{ma07b}, they provide clues about the nature of the true behavior to be expected in 3D.

 \section{2D Simulations of Multiple Simultaneously Burning Shells}
 
 \subsection{Starting Point}
 The initial conditions are a fundamental problem for numerical simulations which explore the development of instabilities to large amplitude. Subtle biases in
 the initial state might make the subsequent development misleading. Use of
 a 1D initial model, with no reliable description of the turbulent velocity field or the
 extent and position of the boundaries of the convection, is a cause for worry, but is
 the best we can do at present.
 
The 2D simulations started from a model obtained from a sequence by P. A. Young (private communication; \cite{fy08} have evolved this model to core collapse
in 1D, and to 3D explosions).
 The initial state was a 1D model of a 23 \sol\ star, mapped to 2D; it is at the latest
 stages of evolution, about one hour prior to core collapse.
 Turbulent convection developed from very small perturbations ($\sim 10^{-3}$ or less, due to mapping onto a different grid) in the unstable regions. The convection rapidly evolved to a
 dynamic state with far larger fluctuations, independent of the initial perturbations.
 The model was 
 similar to that used in \cite{ma07b}, except the computational volume is deeper,
 reaching down past the Si burning shell,  the aspect ratio is larger (a full quadrant
 was calculated),  the abundance gradients were smoother (more like a diffusion
 approximation for convection; see \S~2.1), and the onset of core collapse was much nearer ($\sim1$ hour).

\begin{figure}
\figurenum{1}
\includegraphics[angle=-90,scale=0.5]{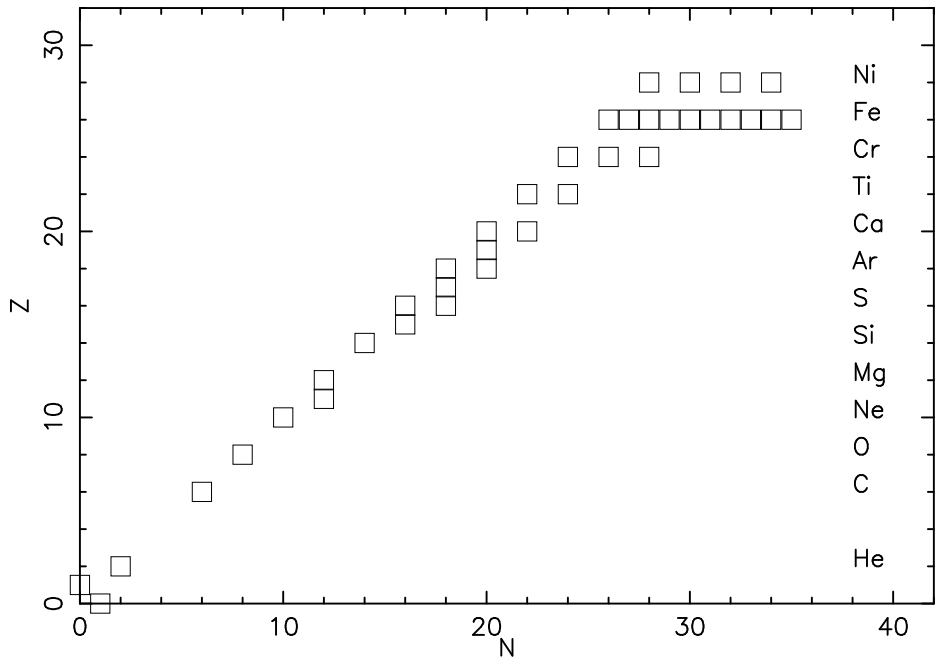}
\caption{Thermonuclear reaction network (37 nuclei) used for burning in C, Ne, O, and Si shells. Each box represents a nucleus; see text for details.}
\label{fig0}
\end{figure}

 The computational domain had an inner boundary representing
 the Fe core and extended well beyond the active burning regions to the
 edge of the He core. The boundary conditions in angle were periodic and
 in radius were reflecting, as in \cite{ma06,ma07a,ma07b}. 
 The equation of state was that of \cite{ts00}, which
 accurately describes the effects of partial relativistic degeneracy of electrons
 and of the formation of electron-positron pairs, and is very similar \citep{ta99}
 to the equation of state used in previous simulations  listed above.
 
 Figure~\ref{fig0} shows the nuclear reaction network used, which contained 38 species (37 isotopes plus electrons).
To deal with increasing neutron excess it was extended to $\rm Ni^{62}$, which 
corresponds to ${\rm Z/A} \sim 0.45$. During Si burning, the nuclei having $\rm Z \ge 22$ approach a local nuclear statistical equilibrium (they become a quasi-equilibrium group). The results of this network were compared to
 those of a 177 isotope network; it reproduced both the energy generation and
 the increase in neutron excess predicted by the larger network (see discussion
of nucleosynthesis and of increase in neutron excess during silicon burning in \cite{wda96}). The nuclear burning was directly coupled to the fluid flow by the method of operator splitting.
 
 \subsection{Results}
 
\begin{figure}
\figurenum{2}
\plotone{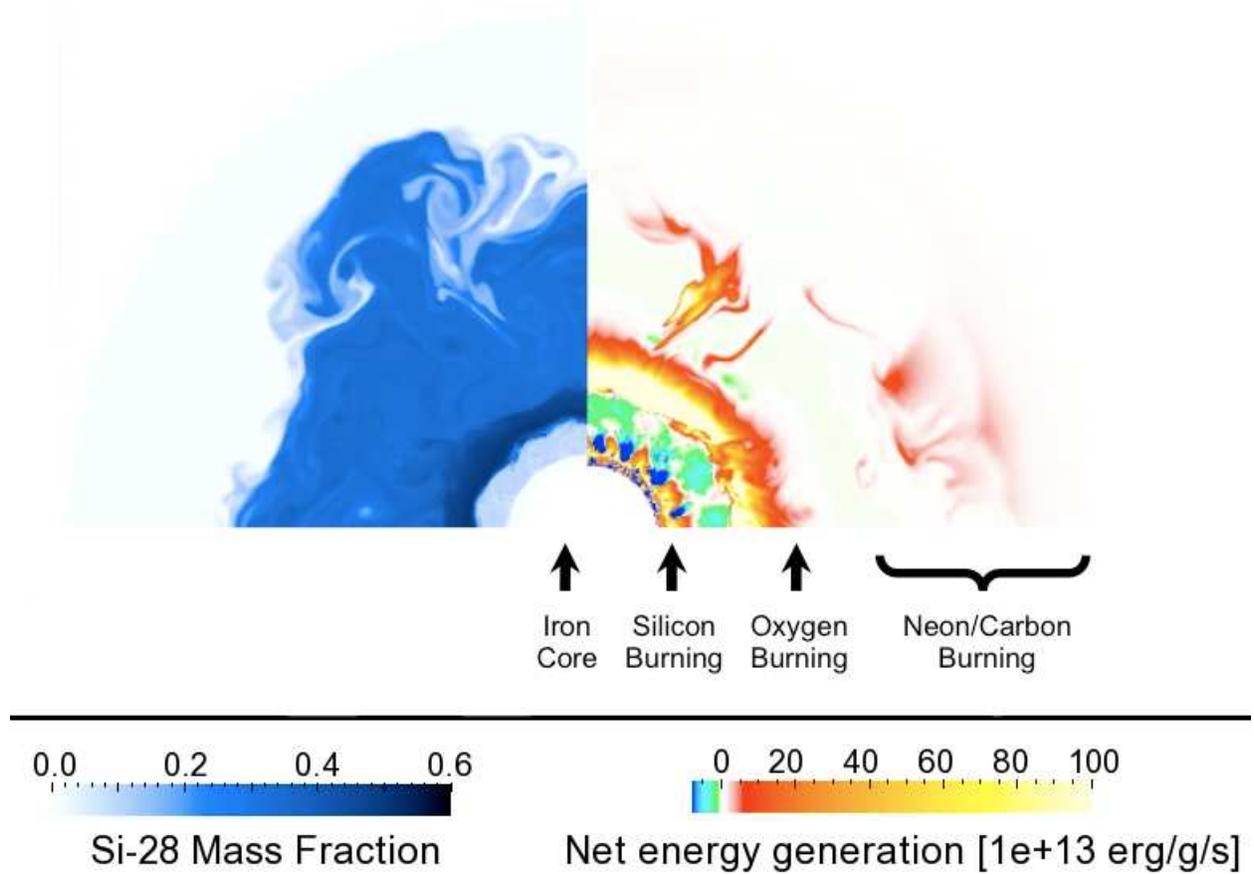}
\caption{A snapshot of the structure of C, Ne, O, and Si shells surrounding the Fe-core of a pre-collapse progenitor of 23 \sol\  star, about 500 seconds after the constraint of spherical symmetry has been removed.  The left side shows the abundance of $\rm Si^{28}$ while the right side shows the net energy generation rate.}
\label{fig1}
\end{figure}

Figure~\ref{fig1} shows the structure of a quadrant of the core of the $23 \msol$ star, with
an iron core (shown as a white semi-circle) in the center.  The computational
domain includes burning shells of Si, O, Ne and C, in order of increasing radius.
The left quadrant displays abundance contours of $\rm Si^{28}$ (dark blue is
high abundance, white is low). The inner light region is a burning shell
which is depleting its Si. Above this is a dark ring which is unburned Si.
Further out is a medium blue layer which contains the O burning shell,
in which Si is being produced. At the top  of this layer are seen streams of
very light blue, denoting entrainment of matter which has not been contaminated by oxygen burning (mostly C and Ne). Finally
there is a very light blue layer from which the streams came, and which
has no enhancement of Si above its original value. 

The right quadrant shows contours of energy generation rate, in units of $10^{13}$
erg/s. Note that the second quadrant is presented as a rotation about the vertical axis; this helps identify corresponding matter in the two different variables
which are mirror images.
The inner ring is again the Si burning shell. It has both strong heating (yellow)
and strong cooling (blue) at the same radius, that is, the burning does not possess
spherical symmetry.  This is enclosed by a green ring, the Si rich layer, which has
milder neutrino cooling, and is no longer spherically symmetric because of ``hot
spots" of burning in entrained (descending) plumes. Beyond this is a ring
of red and yellow which is highly dynamic: the O burning shell itself. 
Finally, above this wisps and plumes of heating are beginning to be seen; these are due to C and Ne burning in entrained matter which is rich in these fuels.

These models have a low Damk\"ohler number \citep{damk},
$D_a = \tau_{turbulence}/\tau_{react} \ll 1$, where the time scale for turbulent flow
is $\tau_{turbulence}$ and the reaction time is $\tau_{react}$. In this limit
  a complex mixing model is not needed to describe the burning, unlike
   burning fronts in type Ia supernova models which require a more complicated
  description.
 Here fuel is transported into regions of higher pressure,
compresses and heats, burns, expands and cools, and is buoyantly transported back to lower pressure. Changes in composition due to turbulent mixing are much faster than those due to nuclear burning, so that a sub-grid flame model is not required.

\begin{figure}
\figurenum{3}
\plotone{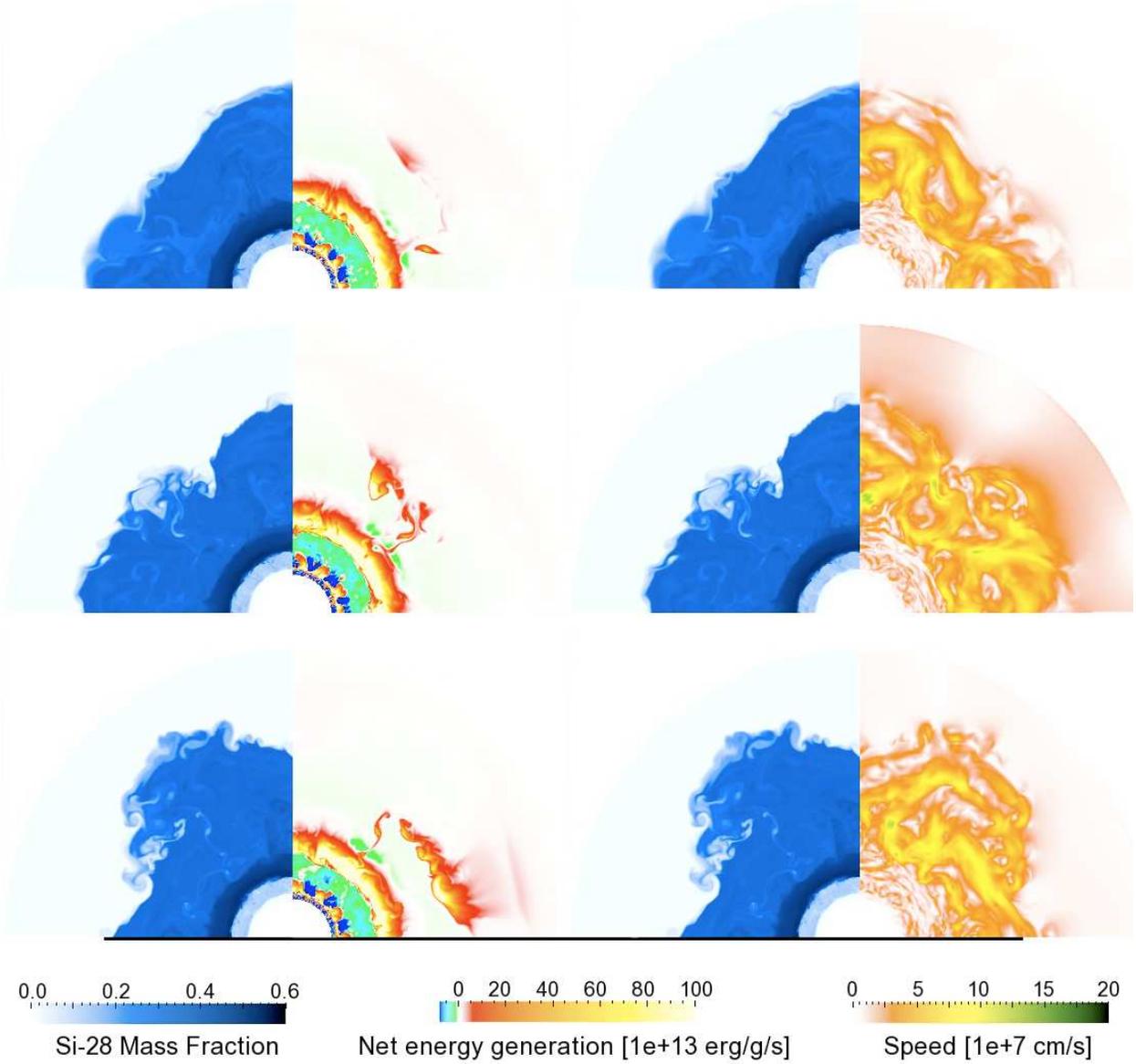}
\caption{Snapshots of the structure of C, Ne, O, and Si shells surrounding the Fe-core of
 a pre-collapse progenitor of 23 \sol\  star.  Three different times are shown,
  $t_f$ = 0, 61, 83 seconds (from top, 0\ s, to bottom, 83\ s)
 after our fiducial model (see text). The left panels (blue) show abundance of
 $\rm Si^{28}$, while the right panels show energy generation rate and
 convective speed, respectively.}
\label{fig2}
\end{figure}
\placefigure{3}

The initial model was strictly spherically symmetric, and had to develop a self-consistent turbulent velocity field to carry the heat from nuclear burning away from
the burning regions. 
During this mild transient phase, asymmetries begin to develop slowly. Figure~\ref{fig1} shows the structure after this development, and at the beginning of significant
deviation from the usually assumed spherical symmetry. Because of the time
needed for the initial spherical model to develop a realistic and consistent convective flux, there is ambiguity regarding the ``initial time"; we simply define
a ``fiducial time"  ($t_f = t-t_0$ where $t_0 \sim 345$ seconds into the 2D simulations) at which the model is still fairly spherical but has a realistic convective flux, and we count elapsed time after that.

Figure~\ref{fig2} shows three snapshots of the structure at different times after the
fiducial time: $t_f$ = 0, 61, and 83 seconds. The left column represents the same variables as shown in Figure~1. The right column shows the $\rm Si^{28}$ abundance as before in the left panel, but in the right panel the energy generation rate is replaced by the turbulent speed in units of $10^7$ cm/s.

Distortions in the O burning shell are obvious. The equation of state in this
region is affected strongly by the thermal production of an equilibrium abundance
of electron-positron pairs, so that the effective adiabatic exponent drops below\footnote{See Table~5 in \cite{amy09}, and recall that
$\nabla_{ad} = (\Gamma_2-1)/\Gamma_2 \approx 1/4$ across the whole O burning convective zone.} 
$\Gamma_1 \sim 4/3$.  Similarly $\Gamma_4 \equiv 1 + E/PV \sim 4/3$;
This means the local contribution to the gravitational binding energy, which is proportional to $\Gamma_4 - 4/3$, is small.
This is a common property of oxygen burning in stars \cite{wda68,wda96}.  The restoring force for stable stratification is weak, allowing large amplitude distortions with little energy cost. 

The decrease in $\Gamma_1$ and $\Gamma_4$ are due to the need to provide
the rest mass for newly formed electron-positron pairs. At low temperature,
$kT \ll 2 m_e c^2$, the number density of pairs relative to the charge density
of ions decreases exponentially with decreasing temperature, and is negligible.
As temperature approaches $ T \sim 2 \times 10^9 $ degrees Kelvin, the
effect on the equation of state becomes largest. At higher temperatures,
the increase in mass of pairs is less relative to the thermal energy $kT$,
so that these gammas approach that of an extreme relativistic gas,
$\Gamma_1 = \Gamma_4 = 4/3$. Oxygen burning ($\rm O^{16}$ fusion) in stars
occurs at $T \sim 2 \times 10^9$ degrees Kelvin, so that this burning stage is
most influenced by the effects of electron-positron pair production on the equation of state.

Consider first the left column in Figure~\ref{fig2}. The top panel (t=0) is relatively
symmetric, but as time passes, the middle and lower panel show increasing
distortion, especially visible at the interface between the outer, light blue layer
and the middle, medium blue layer.
The streams of light blue inside
the medium blue represents entrainment of matter with little Si, that is, C and
Ne fuel. This corresponds to the flame structures seen in the right side of the
left column,
which are due to C and Ne burning (note similarity in shapes on left and right sides in the left column).  
Similar entrainment is occurring at the top of the Si burning convective region;
the outer edge of the light blue inner ring is rippled due to bursts of burning. The amplitude
of these variations is smaller due to the stiffer equation of state here.

The right side of the right column shows the turbulent convective speed. The large
structures are the oxygen burning convective zone. A smaller convection zone
may be seen surrounding the Fe core, due to the Si burning shell. The C-Ne layer,
lying outside the oxygen burning shell, illustrated the effect of a low-order mode.
In the top and bottom panels, there is little motion, while in the middle panel
the amplitude of the motion is near maximum. The lighter red areas at
about 30 degrees and 70 degrees from vertical correspond to nodes in the
modal velocity. Because of symmetry about the equator (horizontal) we have
four nodes in 180 degrees, or an $\ell = 4$ mode being dominant. Odd values
of $\ell$ are suppressed by the domain size and symmetry imposed by our boundary condition, so this is the lowest order possible in this simulation;
it has a period of about 60 seconds, but is mixed with other, weaker modes.
A movie of the simulation shows a dramatic change as this mode turns the speed on and off as we move from top to middle to bottom.

\begin{figure}
\figurenum{4}
\plotone{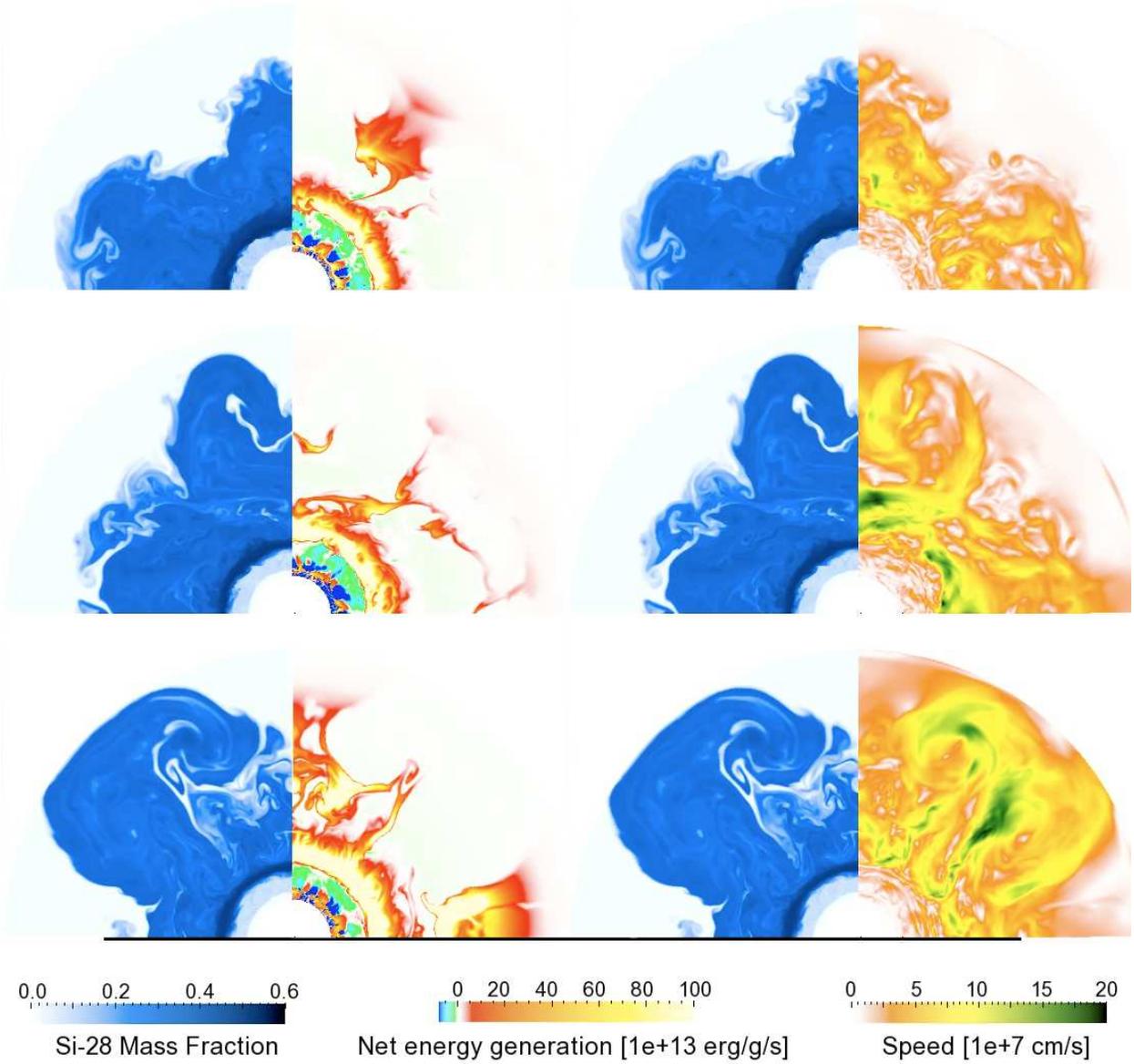}
\caption{Structure of C, Ne, O, and Si shells surrounding the Fe-core of
 a pre-collapse progenitor of 23 \sol\  star. Snapshots at times $ t_f$ = 115, 247, and 307 seconds (top to bottom) after our fiducial model (see text). The format is the same as in Figure~\ref{fig2}. The eruption has become strongly non-linear, as the bottom panels show.}
\label{fig3}
\end{figure}
\placefigure{4}

Figure~\ref{fig3} shows the same variables at  t = 115, 247, and 307 seconds, after several, increasingly vigorous ``sloshes". 
The distortion in the O shell has grown, with large amplitude motions
whose wave-form is
characteristic of $g$-modes. Entrainment  of C and Ne increases.
Fluctuations in Si burning become more vigorous and distort the Si layer
although quasi-equilibrium damps explosive excursions. 
The Si shell burning occurs in dynamically forming and disrupting cells,
qualitatively similar in nature to the 3D cell studied in \cite{am11b}.
A pink tinge indicates weak but widespread burning of
carbon in the outer layers of the computational domain.

At $t_f = 307$ seconds what is best described as an ``eruption" is occurring.
The distortion in the O shell (medium blue bulge) continues to grow; 
it is due to strong wave motion, powered by the nuclear burning.
In the bottom panel, the thickness of the O shell convective layer varies by more than a factor of three as a function of angle; it is thin at the equator and thick at a 45 degree angle. This corresponds to an eruption at that angle. 
Entrainment is correspondingly increased, with consequent burning. 
Waves "slosh" back and forth. The
computational domain and boundary conditions restrict the lowest order modes.
The Si burning shell
is affected by the behavior of the O shell.
There  appears to be no evidence of a slacking in growth of the dynamic
behavior, and extensive mixing is occurring. We ended the simulation because the
extent of mixing is so great that it reaches the grid boundaries, and the
simulation domain becomes inadequate.
The assumption of spherically symmetric structure has clearly failed,
as has the assumption of quasi-hydrostatic structure. 

\section{Implications}

Figures~3-4 show the breakdown of the assumption of spherical symmetry, 
which has been the basis of supernova progenitor models, and much of the interpretation of observational data of supernovae and young supernova remnants.  The $\ell = 4$ spherical harmonic is dominant because the computational domain had only one quadrant and periodic boundaries; a full hemisphere would have allowed the $\ell = 1$ mode to appear.

These simulations, of a stage only one hour before core collapse, show more violent behavior than previous 2D simulations of multiple (CNeO) shells
\citep{ma06}, or 3D simulations with a single burning shell \citep{ma07b}. They become more violent in less time; compare durations
in Table~1.
There are several reasons for this behavior.  (1) This stage is late, only an hour
prior to collapse. (2) The 3D O shell simulations had a
smaller computational domain, so that low order modes were restricted by periodic
boundary conditions. (3) The difference between the 2D three shell (CNeO) and four
shell (CNeOSi) simulations seems to be due to an additional interaction between the O and the
Si shells, which are both active. 

Entrainment of new fuel causes mixing, which is countered by heating from burning, which changes entropy deficits in down-flows to entropy excesses, and halts the down-flow motion. This gives a
natural layering mechanism for the highly dynamic system, a {\em dynamic layering}. 
Obviously the compositional structure, and
the predicted yields, depend upon this effect, which has been ignored (or treated
in a diffusion rather than advection algorithm) in all
1D models of progenitor evolution.
Inter-zonal mixing also would have an impact on yields.

Strong wave generation is observed.  Such waves may become compressional (mixed mode, \cite{ma06}) as they propagate 
into the strong density gradient. The waves will dissipate in non-convective regions, causing heating and slow mixing there, and to the extent that the wave heating is faster than radiative diffusion (which is very slow), expansion will occur. These effects are large enough to be seen in the simulations, but need further
quantification to determine their relative importance.

The Fe core (here a static boundary condition)
 will contract, giving increasingly dynamic behavior. Such a dynamic approach to core-collapse has not been
investigated in multi-dimensional, multi-shell simulations, but may have interesting consequences (see below). The duration of the 2D  CNeOSi simulation was $\sim 600$ seconds, 
while the time for core collapse, neutrino trapping, 
and rebound shock are together a few seconds \citep{wda96}. The observed diffusion time for neutrinos from SN1987A was comparable, also a few seconds. Because time for core collapse is fast in comparison to the period associated with O shell dynamics, the shell structure may be caught in a distorted state by the explosion shock, {\em giving a non-spherical remnant 
even if the explosion shock were perfectly spherical} (which it is unlikely to be;
\cite{kif03,kif06}).

The 2D simulations show much more active dynamical behavior than suggested
by linear perturbation analysis \citep{mbh06}. In \S\ref{time-dep} this was
 traced back to the treatment of turbulent convection in the linear
analysis.  \cite{lorenz} showed that convection has an intrinsic nonlinear instability;
it arises from advection, which gives product terms (XY and XZ in his
notation); these are products of the velocity amplitude with the horizontal
and vertical temperature fluctuations respectively \citep{am11b}.
This provides an explicit example in which
linear perturbative methods for pulsations  \citep{cox80,unno89} require modification when turbulent convection is present. Study of low order dynamical
models, such as  \cite{lorenz}, will provide insight into the nature of this general problem.

It is unlikely that 1D progenitor models are realistic;
in addition to unlikely {\em geometry}, they ignore the vigorous {\em dynamic
behavior}, which becomes manifest in 2D and 3D simulations, in which it is not forbidden as it is in 1D. There may be eruptions prior to core
collapse, there will be large amplitude distortions away from spherical, ``onion''-skin
structure, and there may be modifications of the supernova shock by explosive oxygen burning. Not only can explosion shocks be non-spherical, but the progenitor mass
they propagate through can be asymmetric as well.
The distortions in the progenitor and those induced by the explosion shock
may leave an imprint on the abundances in the supernova and its ejecta.
Of course, rotation will have its own effects which we will have to disentangle.

\section{Some Issues to Reconsider}

\noindent{\bf Progenitor Structure and Fall-back.} 
Mechanisms of explosion and fall-back are all predicated upon the validity of the detailed structure of the 1D progenitor models. 
The old problem of non-explosion of supernova models \citep{wda96} has been alleviated by multi-dimensional simulations of collapsing cores (e.g., \cite{bldom06,bmh09,wjm10}). More
realistic initial models will bring further changes.
For example, the explosion 
and remnant formation could be modified if an eruption occurred prior to and during core collapse. Similarly, expansion of the mantle surrounding the Fe core
could be caused by dynamic burning such as that illustrated in Figures~3-4. 
It would reduce the ram pressure of infall and the mass to be
photo-dissociated, making it easier to eject matter for a given core explosion mechanism. See \cite{bbb82,bethe90,wda96}.

Changes in the mass and entropy of the collapsing core will affect its dynamics.
It has long been apparent that rotational effects too should play a role 
\citep{fh46,fh55}. The historical focus on non-rotating collapse was merely because the non-rotating problem was feasible with computing resources then available. All of these characteristics of the progenitor model might be modified significantly by the use of 3D initial models.
The asymmetry and the rotational structure of the progenitor can be significantly
modified by burning shell interactions, as can the rate of fallback, which depends upon the mantle structure and rotation around the Fe core. See
\cite{fryer99,fbh96,fyh06} for a discussion of  fallback and black hole formation. 

Core collapse is a converging flow (density increases); explosion is a diverging one (density decreases). Asymmetries
in convergent flow grow, which is a problem for inertial-confinement fusion
efforts \citep{lindl}. In divergent slow, asymmetries decrease in importance, and
the flow tends toward the spherically symmetric similarity solutions \citep{sedov}.
Because of the growth of asymmetry during collapse, it is important to have
realistic estimates of the asymmetry in initial models of progenitors; 1D models
are spherical, so that the seeds of asymmetry are introduced numerically or
arbitrarily (e.g., \cite{wjm10,bmh09,nbblo}). 

The 2D simulations above suggest that
progenitors prior to collapse develop large asymmetries in the O shell. What
about the Fe core, which is what collapses? Si burning is active is a layer
of convective cells, so that the asymmetries will tend to average out \citep{am11b}.
Simply scaling from Figure~\ref{fig3} suggests asymmetries in the Si shell of order
of a  few percent or more.  In the absence of simulations including the Fe core
in a dynamical way, we have no plausible quantitative estimates of asymmetries in the Fe core itself prior to collapse.  Turbulence in O and Si shells
will drive fluctuations \citep{am11b}; the resulting motion will induce fluctuations
in the URCA shells in the core, and affect the change of entropy and electron fraction there (see \citep{wda96}, \S11 and \S12). This in turn will affect the mass of the core at collapse, and thus the parameters of the collapse mechanism.
Our understanding of the coupled physics of the dynamics of the core as
it approaches collapse is still uncomfortably vague.

\noindent{\bf Neutron Star Kicks.}
For core collapse progenitors, solitary or in binaries, the internal structure becomes
increasingly inhomogeneous in radial density as evolution procedes, and evolves toward a condensed core and extended mantle structure (Fig.~10.4 in \cite{wda96}).
As the core plus mantle mass of these objects approaches the Chandrasekhar mass from above, this tendency increases. Supernovae of type SNIb and SNIc
have light curves which require that they have such masses just prior to collapse. Consider a simple model in which there is a
point-like core inside an extended mantle. If the core is not located at the center
of mass, then the mantle must be displaced from the center of mass in the opposite
direction. More of the mantle mass lies to one side of the core, so there is a net
gravitational force which pulls the core back toward the center of mass (which does
not move). Similarly the core exerts an equal and opposite pull on the mantle to bring the mantle back toward the center of mass. 
With no dissipation, an oscillation would ensue.
The motion of the core relative to the fluid in the mantle  generates waves 
in the mantle material, providing a means for dissipation, so that in the absence of
driving, the oscillation of the core and mantle about the center of mass would
be damped, and settle to a state in which both the core and mantle are centered
on the center of mass. 

If there were a driving mechanism for core-mantle oscillation, there would be
an asymmetry due to the displacement of core and mantle relative to the
center of mass. The core collapse would give an off-center explosion within the mantle, even if the core collapse gave a perfectly spherical explosion shock
relative to the center of mass of the core. 

Figures~3-4 above suggest that multiple shell burning may provide a driving mechanism for core-mantle oscillation; a computational domain containing an
entire hemisphere would have allowed an $\ell=1$ mode to develop, which is
even more suitable for driving such oscillations. Moreover, there will be a difference in the strength of
driving depending upon the mass of the O burning shell; low mass shells will be 
less effective at driving the heavier core. Slow accretion onto an ONeMg white
dwarf, or evolution to collapse by electron-capture, would occur by O burning in
a low mass shell. However, evolution to core collapse by the instability of a more
massive Fe core generally occurs concurrently with more massive O burning shells, such as described above.  \cite{vdh10} has suggested,
on the basis of data on double neutron stars in the galaxy, that formation of neutron stars from the collapse of ONeMg cores might occur with almost no kick velocity at 
birth, while neutron stars formed by Fe core collapse would receive a large
space velocity at birth. See the review by \cite{kvw08} for background and references.
The discussion above provides a physical mechanism for the empirical
suggestion of van den Heuvel; the size of the kick velocity at collapse will depend
upon the mass of the oxygen shell surrounding the core, and is driven by the
dynamics of multiple shell burning. 

\cite{wjm10} find more vigorous kicks from collapse of Fe cores because the
explosion develops sooner in ONeMg core, and the longer term instability
in the post bounce behavior has less time to be effective.
However, the collapse simulations to date have used small seed perturbations
which may be unrealistic (they are much smaller than those
found in Figures~3-4). 

In addition, the simulations shown in Figures~3-4
assume a {\em static} Fe core, and thus underestimate the total effect:
asymmetry in the Fe core itself should have an
important effect on the collapse and bounce, as mentioned above. 
A relatively
massive mantle may itself affect the post-bounce behavior of the explosion
by setting the initial condition which results in symmetry breaking
(see below), which in the long term gives the hydrodynamical kick \citep{wjm10}.
Either way, the higher kick velocity is associated with a more massive mantle,
which provides the mechanism for symmetry breaking, and a complete
picture needs to be developed.

\noindent{\bf Early $\gamma$-rays.} Gamma rays from the decay of $\rm Ni^{56}$  
were observed in SN1987A before they were expected, based upon 1D progenitor models \citep{abkw89}.  
A  strong O shell eruption, if hot enough to produce some Ni, followed by convective buoyancy and penetrative convection {\em prior to collapse}, would explain the early detection of gamma-rays, with no new hypotheses. 
Alternatively, explosive burning during the
passage of the ejection shock would give an uneven distribution of fresh $\rm Ni^{56}$.
Either way, some $\rm Ni^{56}$ would be moved out further than in a spherically symmetric model, allowing earlier escape of $\gamma$-rays.

\noindent{\bf Young SN Remnants}.
Young SN remnants have not yet mixed with the interstellar medium, and contain
abundance information about the progenitor.
The dynamic nature of pre-collapse evolution adds a new consideration to 
attempts to connect progenitor models to
observations of young supernova remnants, such as Cas~A  \citep{yeafr08}. 
For example, the puzzling inversion of Fe relative to Si found by \cite{hughes00} in Cas~A could easily be explained by vigorous dynamics of the O shell prior to core collapse.
The spherical 1D models are likely to be an inadequate basis for interpreting
observational data (e.g., \cite{fesen88,jds09}), which may now be reanalyzed
with a broader insight. Aspherical shock waves
from the collapsed core become more spherical as they propagate outward
\citep{wjm10}, so that even a very non-spherical collapse may be ineffective at
producing asymmetries in the O shell. However, pre-existing asymmetries in the  O shell, already significant, will be enhanced by explosive burning as the shock passes.

\section{Summary}

The evolution of core-collapse progenitors is likely to be strongly dynamic,
non-spherical, and may have extensive inter-shell mixing. These effects are
ignored in existing progenitor models.

The cyclonic patterns typical of 2D simulations are unstable in 3D, breaking apart and becoming the turbulent cascade of Richardson and Kolmogorov. This enhances damping, and results in the lower velocities seen in 3D relative to 2D simulations. However, simulations in 3D will have essentially the same driving
mechanisms as in 2D. Based upon existing simulations (e.g., \cite{ma07b}), it is unlikely that the increased damping will eliminate dynamic behavior entirely. The increased damping may be able to moderate the eruptions seen in 2D, so that a set of quasi-steady dynamic pulses develops and continues until
the core collapses. Alternatively, the increased damping may be inadequate to prevent continued growth of the instability, so that eruptions such as seen in Figure~\ref{fig3} will develop anyway, at a later time. The ultimate behavior would then be decided well into the non-linear regime. An extreme case would be an explosion
powered by O burning prior to collapse of the Fe core. The observed light curve would depend upon the mass, kinetic energy and amount of ejected $\rm Ni^{56}$
\citep{wda96}.

We need full 3D simulations to determine the quantitative impact of these new phenomena.
From the discussion above it is possible to determine the
features needed for such simulations:
\begin{itemize}
\item {\bf Full $4\pi$ steradians,} including the whole core, to get the lowest order fluid modes ($\ell=1$), rotation, and low order MHD modes,
\item {\bf Real EOS,} to capture effect of electron-positron pairs and relativistic partial degeneracy,
\item {\bf Network,} for realistic burning of C, Ne, O and especially Si with e-capture in a dynamic environment,
\item {\bf Multiple shells,} (C, Ne, O, Si) to get shell interactions,
\item {\bf Sufficient resolution,} to get turbulence
and to calculate coherent structures (ILES), and 
\item {\bf Compressible fluid dynamics,} to get mixed mode waves, and possible  eruptions.
\end{itemize}

Low Mach number solvers such as MAESTRO \citep{maestro} may be useful, if generalized to include a dynamic background (the core evolution accelerates),
or applied to earlier stages of neutrino cooled evolution (which may  be strongly subsonic).

This is a challenging combination of constraints, but such computations are becoming feasible.
If we scale from the \cite{ma07b} 3D simulation, the increased solid angle gives a factor of 85, the increase in radius a factor of 2, for a total increase of 170. A further
increase in the radius of another factor of 2 would increase the computational load
by a factor of 340, and would allow investigation of eruptions further into the
strongly nonlinear regime than shown here.
However that simulation was done on
a small Beowulf cluster ($\sim100$ cpus, which were slower than available now).
This factor of 340 from the increased computational domain is more than balanced
by the increased computational power available with top level machines.  Doing 
an equivalent simulation, but in 3D, is feasible. More difficult is including the
Fe core, which requires a different grid near the origin, but this has already been
solved in different ways by several groups (see \cite{wpj03,dearborn06,wjm10}). 
The computational demands, of a 3D simulation of the evolution of a progenitor into hydrodynamic core collapse, seems to be no worse than the computational demands of a single 3D core collapse calculation through bounce.

\begin{acknowledgements}
This work was supported in part by NSF Grant 0708871,
NASA Grant NNX08AH19G at the University of Arizona, and
by ARC DP1095368 (J. Lattanzio, P. I.) at Monash University, Melbourne, Australia.
Dr. T. Janka, Dr. E. M\"uller, Prof. F. Timmes and Prof. P. A. Young provided
detailed comments and questions which helped us tighten the presentation.
One of us (DA) wishes to thank 
Prof. Remo Ruffini of ICRAnet, and Prof. John Lattanzio
of CSPA, Monash University,
Peter Wood of Australian National University / Mount Stromlo
Observatory, and the
Aspen Center for Physics for their hospitality and support.

\end{acknowledgements}

\eject

\clearpage

\end{document}